\documentclass[preprint,12pt]{elsarticle}
\usepackage{amssymb}
\usepackage{amsmath}
\usepackage{multirow}
\usepackage{bm}
\usepackage{soul}
\usepackage{color}
\usepackage{xcolor} % Required for defining custom colors if needed
\usepackage{hyperref}
\hypersetup{
    colorlinks=true,  % Colours links instead of ugly boxes
    urlcolor=blue,     % Colour for external hyperlinks
    linkcolor=blue,    % Colour of internal links (sections, figures, etc.)
    citecolor=red     % Colour of citations
}

\journal{International Journal of Mechanical Sciences}

\begin{document}

\begin{frontmatter}

\title{Learning interpretable surface elasticity properties from bulk properties}

\author[1]{Saaketh Desai}
\author[1]{Prasad Iyer}
\author[1]{R\'{e}mi Dingreville\corref{r1}}
\cortext[r1]{rdingre@sandia.gov}
\affiliation[1]{
                organization={Center for Integrated Nanotechnologies, Sandia National Laboratories},
                city={Albuquerque},
                state={NM},
                postcode={87123},
                country={US}
}

%% Abstract
\begin{abstract}
Surface elasticity is central to understanding the mechanics and stability of surfaces and interfaces.
It is characterized by quantities such as surface tension, residual surface stress, and surface stiffness, however their analytical expressions are typically difficult to derive from atomistic data, and depend strongly on modeling choices.
This work presents a neural network-based equation learner which combines customized activation functions and connection-based pruning to discover parsimonious, closed-form equations for surface elasticity from atomistic simulations.
Applying the method to seven face centered cubic (FCC) metals, our equation learner uncovers interpretable equations that describe both low-Miller index and high-Miller index surface properties, capturing long-tail property distributions accurately.
The discovered expressions are decoupled into two components: a universal, geometry-driven orientation function, and material-specific baseline coefficients.
We find that lower-order properties such as surface tension are fundamentally geometry dependent, while higher-order properties such as surface stress and elasticity show more complex geometry and material dependence.
We also relate material dependent coefficients to bulk properties, forming a clear map from bulk material properties to surface elasticity.
Overall, this approach demonstrates that interpretable neurosymbolic machine learning can bridge the gap between atomistic simulations and physical laws, enabling the discovery of generalizable structure-property relationships for materials science phenomena such as surface elasticity.
\end{abstract}

%%Graphical abstract (optional, but encouraged)

% Highlights
\begin{highlights}
\item Neuro-symbolic regression discovers interpretable closed-forms for surface elasticity
\item Surface properties decoupled into geometry and material-specific parts
\item Learned equations directly map bulk properties to surface elasticity 
\end{highlights}
% Keywords (required)

\end{frontmatter}
\pagebreak

%%%%%%%%%%%%%%%%%%%%%%%%%%%%%%%%%%%%%%%%%%%%%%%%%%%%%%%%%%%%%%
%% INTRODUCTION
%%%%%%%%%%%%%%%%%%%%%%%%%%%%%%%%%%%%%%%%%%%%%%%%%%%%%%%%%%%%%%
\section{Introduction}\label{sec:intro}
\par{
Surface elasticity is a central concept in surface science describing the mechanical behavior and stability of surfaces and interfaces.
Surface elasticity is particularly important for nanoscale systems, where surface effects become prominent and dominate over bulk properties due to the high surface-to-volume ratio~\cite{miller2000size, sander2003surface, sharma2003effect, dingreville2005surface, ansari2011surface, li2014dependence, hu2021singular, pan2025surface, zhu2025surface}.
For solid surfaces, a key concept is the distinction between surface tension ($\Gamma$) and surface stress ($\Sigma^S$).
While both terms are often used interchangeably, especially in the context of liquids, they represent different physical quantities for solid surfaces.
Surface tension (also referred to as the Gibbs surface excess energy) is a scalar quantity that represents the energy required to create new surface area, while surface stress is a second-rank tensor that describes the forces within the surface necessary to reversibly and elastically deform it~\cite{dingreville2005surface, kramer2007note}.
For solid surfaces, the relationship between $\Gamma$  and $\Sigma^S$ is described by the Shuttleworth-Herring equation~\cite{Shuttleworth50, Herring51} which expresses $\Sigma^S$ as the variation of $\Gamma$ with respect to the in-plane surface deformation, $\epsilon^S$, such that $\Sigma^S = \partial\Gamma/\partial \epsilon^S$.
The surface elasticity components describing the constitutive relationship between $\Sigma^S$ and $\epsilon^S$ consist of
a residual surface stress tensor, $\Sigma^{S,0}$, resulting from the relaxed configuration of the surface, and
a surface elasticity stiffness tensor, $\mathbb{C}^{S}$, a fourth-rank tensor,describing the anisotropic elastic response to surface to deformations~\cite{Gurtin75, dingreville2005surface}.
These surface tensors can also be expressed in terms of a set invariants, allowing for direct comparison of surface properties across different surface orientations~\cite{chen2022invariant}.
While $\Gamma$ must be positive for all stable surfaces, the components of $\Sigma^{S,0}$ and $\mathbb{C}^{S}$ can be positive or negative and depend on the material being considered and the surface orientation~\cite{dingreville2007semi}.
}
\par{
Because of their importance for so many properties in nanomaterials~\cite{park2008surface, ansari2011bending, gheshlaghi2011surface, wang2011surface}, nanoscale fabrication processes~\cite{ibach1997role, fischer2008role}, and capillary problems~\cite{wang2007effects, andreotti2016soft, xu2017direct}, accurate computation of surface elastic properties are essential to correctly describe the thermodynamics of solid surfaces and predict the overall behavior and stability of nanostructured materials.
Measuring or calculating surface elastic properties presents significant challenges.  
Factor like surface relaxation, anisotropy, orientation of the surface can all affect the surface elastic properties.
Experimental techniques such as contact-angle measurements~\cite{tyson1977surface} can determine surface tension, but measuring the residual surface stress or surface stiffness is a lot more challenging, often relying on indirect and approximate methods like the zero-creep method~\cite{digilov1976}.
From a theoretical point of view, surface elastic properties are typically calculated using atomistic simulations such as density functional theory~\cite{needs1991theory} or molecular dynamics calculations~\cite{shenoy2005atomistic, dingreville2007semi,dingreville2009semi}.
However, such calculations are limited to low Miller index planes and results vary depending on the interatomic potential and surface relaxation configuration used~\cite{dingreville2007semi}, precluding a universal description of surface properties across arbitrary orientations and materials.
}

\par{
Machine-learning methods offer an alternative solution to accelerate, improve, and expand our ability to calculate surface elastic properties~\cite{ye2021deep, zhang2021machine, chen2022invariant}.
For instance, Khoei et. al. \cite{khoei2024machine} used machine learning to directly predict the non-linear mechanical response of nanocrystalline structures.
More directly, Chen and coworkers~\cite{chen2022invariant} used neural networks and boosted regression trees to predict invariant surface elastic properties for seven face-centered cubic (FCC) metals (Ag, Al, Au, Cu, Ni, Pd, and Pt), using  the regression tree predictions to explore potential links between surface and bulk elastic properties.
However, such `black-box' machine-learning approaches limit our ability to provide an interpretable link between bulk properties, surface orientations, and surface elastic properties, which restricts their effectiveness in uncovering physical mechanisms at play.
Additionally, caution is required when applying these models out-of-distribution, as they often struggle to extrapolate accurately to new materials.
}

\par{
In recent years, extracting interpretable equations from data has gained attention, with early methods such as symbolic regression~\cite{schmidt2009distilling} using genetic algorithms to learn equations from tabular data, and methods such as sparse regression~\cite{brunton2016discovering} for discovering partial differential equations (PDEs) from data.
More recent efforts have shown the promise of neural networks for equation learning (nn-EQL), as seen in customized networks developed by Sahoo et al.~\cite{sahoo2018learning} and Desai et al.~\cite{desai2021parsimonious}, where specialized activation functions, and genetic algorithms~\cite{desai2021parsimonious} are used to favor the discovery of parsimonious equations.
Udrescu's AI-Feynman~\cite{udrescu2020ai} and subsequent AI-Descartes~\cite{cornelio2023combining} offer more holistic frameworks for learning and deriving equations, but have been limited to rediscovering known equations.
Variants of symbolic regression such as sparse regression have also shown some success in learning constitutive laws of materials \cite{im2022discovering}.
A key challenge in these equation learning models is parameter redundancy, where multiple parameters may contribute redundantly to the model's predictions.
Conventional regression models such as neural networks do not inherently prioritize simpler solutions, and while their large size enables them to capture highly complex patterns in the data, potential redundancy can also lead to overfitting.
Specifically for learning equations, this parameter redundancy can result in `bloated' equations with too many parameters.
Therefore, equation learning approaches use various regularization or architectural constraints in an attempt to discover the simplest form of an equation.
%
%However, these regularization or architectural constraints have varying degrees of success, and may not scale to larger models.
%
One set of regularization strategies that has proven successful in conventional deep learning is pruning.
Pruning strategies aim to reduce complex neural networks by identifying smaller networks that still perform well without overfitting~\cite{lecun1989optimal, hassibi1992second}.
Strategies such as iterative magnitude pruning~\cite{frankle2018lottery} and soft-thresholding~\cite{kusupati2020soft}, have been quite effective in deep networks at scale.
These approaches remove redundant weights entirely, as opposed to penalizing their magnitude via terms in a loss function, as is typical in regularization approaches.
However, these methods have yet to be widely applied to equation learning, where simplification would greatly enhance interpretability.
}

\par{
In this paper, we are particularly interested in nn-EQL approaches to learn equations that describe surface elasticity from bulk properties, directly from data.
We build on both the previously developed database from Chen et al.~\cite{chen2022invariant} and advances in nn-EQL~\cite{desai2021parsimonious} to learn interpretable relationships describing surface elasticity.
In what follows, Section 2 describes the methods used to compute surface elastic properties, and the nn-EQL workflow.
Section 3 demonstrates that the nn-EQL accurately captures lower- and higher-order surface properties, including long-tail distributions.
Section 4 then describes the equations discovered, uncovering a universal geometry-dependent function, and material specific coefficients.
Section 5 then relates the material specific coefficients to bulk properties, creating a map from bulk properties to surface elasticity.
}

%%%%%%%%%%%%%%%%%%%%%%%%%%%%%%%%%%%%%%%%%%%%%%%%%%%%%%%%%%%%%%
%% Methodology
%%%%%%%%%%%%%%%%%%%%%%%%%%%%%%%%%%%%%%%%%%%%%%%%%%%%%%%%%%%%%%

\section{Methods}
\label{sec:methodology}
\subsection{Surface elastic properties}
\par{
The surface elastic properties of interest (surface tension $\Gamma$, surface stress $\Sigma^{S,0}$, and surface elasticity tensor $\mathbb{C}^S$) are represented by their invariants considering plane elasticity within a polar method~\cite{vannucci2005plane,chen2022invariant}.
Since the surface tension $\Gamma$ is a scalar, it is already in an invariant form.
The second-rank residual surface stress tensor, $\Sigma^{S,0}$, can be represented by two rotational invariants, $T$ and $R$, and a polar angle, $\Phi$ such that,
\begin{equation}
\begin{aligned}
    \Sigma^{S,0}_{11} & = & T + R\cos2\Phi,\\
    \Sigma^{S,0}_{22} & = & T - R\cos2\Phi,\\
    \Sigma^{S,0}_{12} & = & R\sin2\Phi.
\end{aligned}
\end{equation}
$T$ and $R$ represent the spherical and deviatoric parts of the residual surface stress respectively.
The polar angle $\Phi$ represents the rotation about the axis normal to the surface.
For instance, for the isotropic surface $\langle100\rangle$, $\Sigma^{S,0}_{11} = T$ (assuming that due to the surface symmetry, $R$ is null, see Appendix B in Ref\@.~\cite{chen2022invariant}).
The fourth-rank surface elasticity stiffness tensor, $\mathbb{C}^S$, can be represented by four invariants, $T_0, T_1, R_0$, and $R_1$, and two polar angles, $\Phi_0$ and $\Phi_1$ such that,
\begin{equation}
\begin{aligned}
    \mathbb{C}^S_{1111} & = & 2T_1 + T_0 + 4R_1\cos2\Phi_1 + R_0\cos4\Phi_0,\\
    \mathbb{C}^S_{2222} & = & 2T_1 + T_0 - 4R_1\cos2\Phi_1 + R_0\cos4\Phi_0,\\
    \mathbb{C}^S_{1122} & = & 2T_1 - T_0 - R_0\cos4\Phi_0,\\
    \mathbb{C}^S_{1212} & = & T_0 - R_0\cos4\Phi_0,\\
    \mathbb{C}^S_{1112} & = & 2R_1\sin2\Phi_1 + R_0\sin4\Phi_0,\\
    \mathbb{C}^S_{2212} & = & 2R_1\sin2\Phi_1 - R_0\sin4\Phi_0.
\end{aligned}
\end{equation}
$T_0$ and $ T_1$ represent the isotropic part of the surface stiffness tensor, while $R_0$ and $R_1$ correspond to the anisotropic parts.
For instance, for the isotropic surface $\langle100\rangle$, $\mathbb{C}^S_{1111} = 2T_1+T_0$ (assuming that due to the surface symmetry, $R_1$ and $R_0$ are null, see Appendix B in Ref\@.~\cite{chen2022invariant}).
Note that the two polar angles $\Phi_1$ and $\Phi_0$ are rotation angles from two distinct reference frames associated with the surface stiffness tensor.
However, from the polar method, $\Phi_1-\Phi_0$ is set as a constant value equal to K$\pi/4$, with K = 0 or 1 (see Eq\@.~(12) in Ref\@.~\cite{chen2022invariant}), reducing the number of invariants necessary to describe $\mathbb{C}^S$ to only five invariants.
We refer the reader to Refs.~\cite{vannucci2005plane} and~\cite{chen2022invariant} for a complete description of those surface elasticity invariants.
}

\par{
We used the calculated surface invariants defined above from prior work ~\cite{chen2022invariant}.
These invariants were generated for seven FCC metals (Ag, Al, Au, Cu, Ni, Pd, and Pt) via molecular statics simulations using the molecular dynamics simulation code LAMMPS (Large-scale Atomic/Molecular Massively Parallel Simulator)~\cite{thompsonLAMMPSFlexible2022}.
The interatomic potentials used to describe these metals include the potential developed by Foiles et al.~\cite{foiles86embedded} for Ag, Au, Cu, Ni, Pd, and Pt and Jacobsen et al.~\cite{jacobsen87interatomic} for Al.
These potentials were fit to bulk elastic constants, cohesive energy, and low-index surface energies (\textit{i.e.}, $\Gamma$).
To simulate the free surface configurations, a perfect crystal lattice was rotated to achieve the desired surface orientations within a rectangular simulation cell.
The surface orientation was represented by two spherical coordinates $\theta$ and $\phi$.
The angle $\phi$ is defined in the surface plane ($x_1-x_2$ plane), while the angle $\theta$ is measured from the surface normal, \textit{i.e.}, the $x_3$ axis.
Periodic boundary conditions were applied in the in-plane directions ($x_1-x_2$) to mimic an infinite film, with the film thickness (along the $x_3$-direction) chosen to ensure independence from thickness effects.
The simulation cell had a dimension of $30\times30\times15~\rm{nm}^3$, while a $15\times15\times15~\rm{nm}^3$ sub-cell was used to calculate surface properties from the relaxed free surface, avoiding any boundary effects.
Each free-surface atomistic system was relaxed via conjugate gradient energy minimization.
Surface elastic properties were computed using a semi-analytical atomistic method~\cite{dingreville2007semi} from those relaxed, free-surface configurations.
Bulk properties were calculated from a sub-volume extracted from the center of the simulation cell away from the free surface and the boundaries.
The computed bulk properties include
the material lattice parameter ($a$),
the stacking-fault energy ($E_{\rm{SF}}$),
the cohesive energy ($E_{\rm{coh}}$),
the bulk modulus ($K$),
the $\{001\} \langle110\rangle$ shear modulus ($G' = 1/2(\mathbb{C}_{1111} - \mathbb{C}_{1122}$)),
the $\{001\}\langle110\rangle$ shear modulus ($G'' = \mathbb{C}_{2323}$).
In the above the tensor $\mathbb{C}$ denotes the bulk elastic stiffness tensor.
Further details on the computation of the surface and bulk properties considered here are provided in~\cite{chen2022invariant}.
}

\par
\begin{figure}[h!]
\centering
\includegraphics[width=0.99\linewidth]{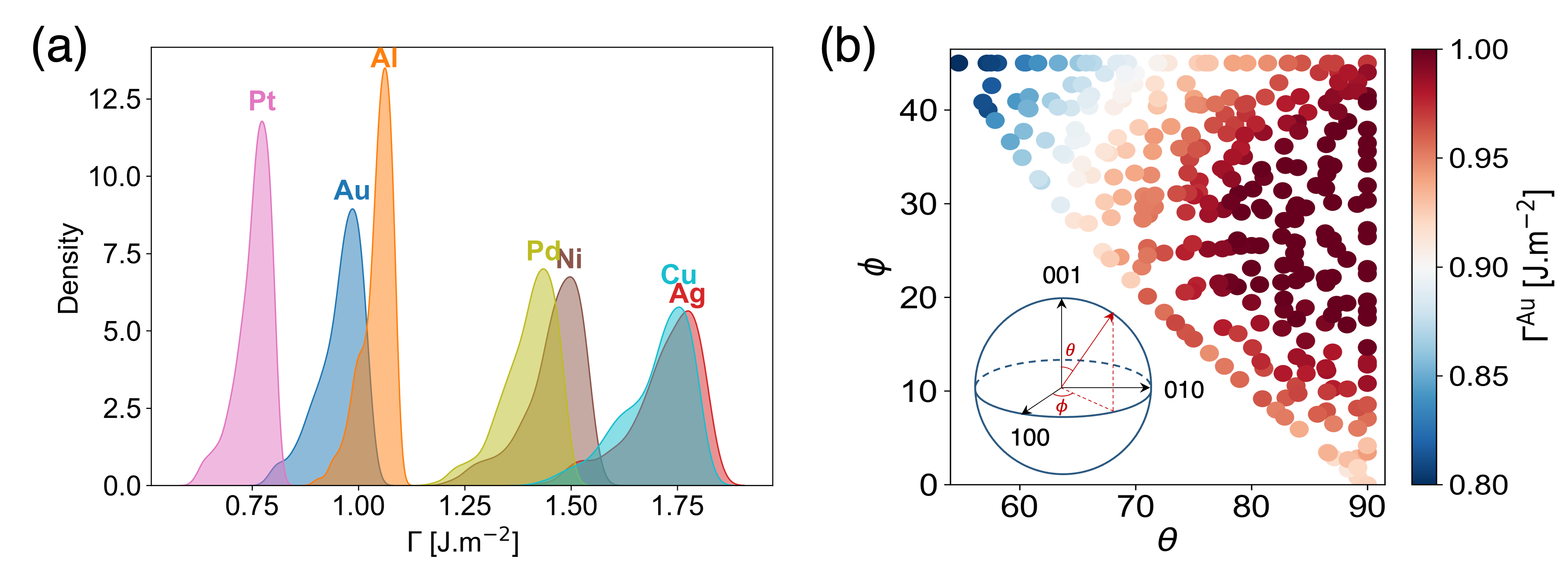}
\caption{\textbf{The dataset of surface properties:}
(a) Distributions of surface tension ($\Gamma$) for seven FCC metals.
(b) Range of surface orientations covered and corresponding surface tension values for Au, displayed in terms of spherical coordinates, $\theta$ and $\phi$.}\label{fig:dataset}
\end{figure}

{
In total, considering the seven FCC metals, the dataset of surface elastic properties consisted of 2,128 different surface configurations, with 304 surface orientations for each material.
The bulk properties were calculated for all seven FCC metals.
Fig\@.~\ref{fig:dataset}(a) shows the distribution of surface tension across all seven metals, while Fig\@.~\ref{fig:dataset}(b) illustrates the range of surface orientations ($\theta$, $\phi$).
}
%%%%%%%%%%%%%%%%%%%%%%%%%%%%%%%%%%%%
%% FIG dataset
%%%%%%%%%%%%%%%%%%%%%%%%%%%%%%%%%%%%
%%
%%%%%%%%%%%%%%%%%%%%%%%%%%%%%%%%%%%%
\subsection{Neural network equation learner (nn-EQL) workflow}\label{sec2}

\par
\begin{figure}[!ht]
\centering
\includegraphics[width=0.99\linewidth]{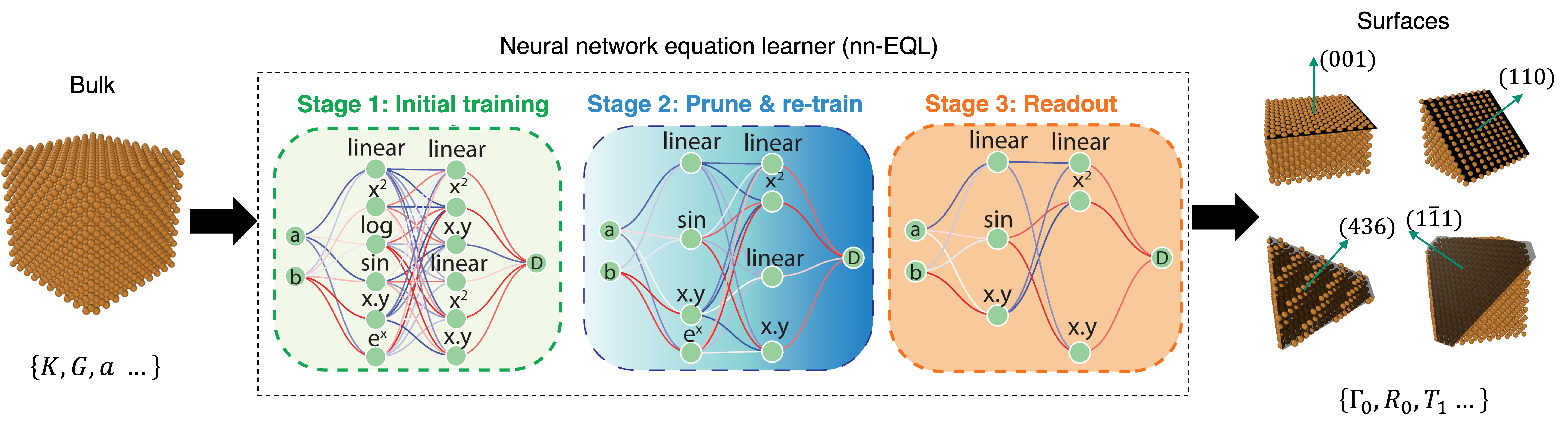}
\caption{\textbf{Equation learner (nn-EQL) workflow:} We learn interpretable relationships between bulk properties of materials (e.g., bulk modulus, shear modulus, lattice parameter) and surface properties of materials (invariants) using equation learner networks. nn-EQLs are trained in three stages:
Stage 1 is the training of an initial network with customized activations.
Stage 2 uses connection-based pruning to reduce network size.
Stage 3 (post training) is a readout of the network weights and activations as an equation.}\label{fig1}
\end{figure}

{
We developed a nn-EQL to learn interpretable relationships for surface elastic properties.
Our nn-EQL architecture in inspired from prior work~\cite{desai2021parsimonious}, using a shallow, two-three layer network with special activation functions, as well as a modified iterative magnitude pruning scheme.
As illustrated in Fig\@.~\ref{fig1}, this nn-EQL takes as input bulk properties and predicts a functional representation of the invariants of the surface elastic properties as an output.
By employing a pruning scheme, the nn-EQL favors parsimony, learning the smallest equation that fits the data well, and avoids redundancies.
}

\par{
We first began by training a basic nn-EQL network using 1-3 layers, where each neuron in a layer has customized activation functions such as $\sin$, $\cos$, $\exp$, and $\log$, along with polynomial activation functions such as $x$, $x\textsuperscript{2}$, $x\textsuperscript{3}$, etc.
Our motivation to use these activation functions is to learn equations with terms typically present in equations in physical sciences.
This network aims only to fit the data well, and can be as large as needed (\textit{i.e.} non parsimonious), as long as the network does not overfit (\textit{i.e.} the error on the validation set is the lowest it can be).
Once this network was trained to a desired accuracy both on the training and validation sets, we began cycles of iterative pruning to simplify the symbolic expressions.
}

\par{
To simplify the functions learned by this initial nn-EQL, we adopted a pruning strategy which consists of setting the bottom $p$\% of weights to 0 and kept fixed for the rest of the training every $n$ epochs.
In this context, $n$ and $p$ are the pruning hyperparameters to be selected.
In conventional neural network pruning~\cite{frankle2018lottery, kusupati2020soft}, weights with the lowest magnitude are pruned, with varying considerations for smooth gradient flow during back-propagation.
Other parsimony strategies such as genetic algorithms also effective in learning equations, as shown in Ref. \cite{desai2021parsimonious}.
However, our use of customized, non-monotonic activation functions in the nn-EQL implies that removing the smallest weights in each layer may not correspond to removing the smallest connections in the network (\textit{e.g.}, $\cos(wx)$ is large when the weight $w$ is small, for a given $x$).
Therefore, we used a connection-based pruning in our work, where a connection's strength is determined as the product of the weight and the activation of the neuron in the previous layer.
This allows small weights in conjunction with specific activations to still retain presence in the network.
Here, after some calibration tests, we chose $n$ and $p$ to be set to 500 and 10 respectively.
That is, we pruned 10\% of the weights every 500 epochs.
After rounds of connection-based pruning, we aimed to achieve a network that is 1 to 10\% of the original network, resulting in a smaller equation that still fits the data well.
This pruned network typically corresponds to $\sim$10 parameters, resulting in interpretable equations with $\sim$5-7 terms.
We used Pytorch \cite{paszke2019pytorch} to set up the nn-EQL, using Pytorch pruning to efficiently and automatically prune the network.
Finally, after a pruned network is deemed to have achieved sufficient accuracy on the training and validation set, we used the Sympy \cite{meurer2017sympy} library to automatically convert the network's weights and activation functions into a human-readable equation.
}
%%%%%%%%%%%%%%%%%%%%%%%%%%%%%%%%%%%%
%% FIG workflow
%%%%%%%%%%%%%%%%%%%%%%%%%%%%%%%%%%%%
%%
%%%%%%%%%%%%%%%%%%%%%%%%%%%%%%%%%%%%
\par{
As listed in Table~\ref{tab:eql_benchmark}, to validate our equation learner network,  we attempted to rediscover a set of known equations used to benchmark other intepretable ML workflows, such as AI-Feynman \cite{udrescu2020ai}, or other symbolic regression workflows \cite{brunton2016discovering}.
Ref. \cite{desai2025autoscilab} documents this benchmarking, comparing true and discovered equations.
We find that the discovered equations are close to the true equation for examples representing some of  the Strogatz ordinary diffential equations from \cite{la2021contemporary}, and for examples representing equations from the Feynman lecture series, used in \cite{udrescu2020ai}.
%
%We find that our nn-EQL implementation can exactly (re)discover the ground truth equations.
%
While this benchmark was performed only on three examples, these results show that our nn-EQL implementation can (re)discover simple equations commonly found in physical systems, providing credence to our approach.
}
%%%%%%%%%%%%%%%%%%%%%%%%%%%%%%%%%%%%
%% TABLE 1
%%%%%%%%%%%%%%%%%%%%%%%%%%%%%%%%%%%%
\begin{table}[!ht]
\centering
\caption{Benchmarking the neural network equation learner}
\begin{tabular}{|c|l|}
\hline\hline
\textbf{Data set}   & \textbf{True} / \textcolor{blue}{\textbf{Discovered}} \textbf{equation}\\
\hline
\multirow{2}{*}{Lotka-Volterra interspecies dynamics:} & $\dot{x} = 3x - 2xy -x^2$\\
    & {\color{blue}{$\dot{x} = 3.063x - 2.02xy -x^2 - 0.16$ }}\\
\hline
\multirow{2}{*}{Van der Pol oscillator:} & $\dot{x} = 10(y - \frac{1}{3}(x^3 - x))$\\
& {\color{blue}{$\dot{x} = 10(y - 0.3153(x^3 - 1.03x))$}}\\
\hline
\multirow{2}{*}{Magnetic moment of an electron in an orbit:} & $\mu = qvr$\\
& {\color{blue}{$\mu = q v r$}}\\
\hline
\hline
\end{tabular}
\label{tab:eql_benchmark}
\end{table}
%%

%%%%%%%%%%%%%%%%%%%%%%%%%%%%%%%%%%%%%%%%%%%%%%%%%%%%%%%%%%%
%% RESULTS AND DISCUSSION
%%%%%%%%%%%%%%%%%%%%%%%%%%%%%%%%%%%%%%%%%%%%%%%%%%%%%%%%%%%
\section{Results and discussion}\label{sec:resultsanddiscussion}
\subsection{Model performance and validation}

%\subsection{Accuracy across different materials}
{
We first show that our equation learner network can accurately predict surface energies and surface elastic properties.
Figure~\ref{fig3} (a), (c), and (e) show distributions of the surface tension $\Gamma$, the spherical part of the surface residual stress invariant $T$, and the isotropic component of the surface elasticity stiffness invariant $2 T_1+T_0$, for all seven FCC materials in this study.
The network's predictions closely match those obtained from the semi-analytical method~\cite{dingreville2007semi}, validating the generalization of the nn-EQL across all orientations, and all the materials tested, despite significant inherent differences in their surface energy characteristics and surface elastic responses.
Indeed, we observe that the nn-EQL accurately captures the distribution of both lower- and higher-order surface elastic invariants across all materials studied.
Notably, we observe the ability of the nn-EQL to correctly learn the long-tailed nature of distributions for specific properties in certain metals.
These long tails typically correspond to less stable, or high-index surface orientations, which are essential to know for understanding the stability mechanisms of nanostructured objects~\cite{bach1997stress,ibach1997role,dingreville2005surface}.
For instance, the observation that Pt and Ni develop pronounced tails for all three invariants, while Al only exhibits this behavior for surface stiffness, highlights both material-specific surface physics and the importance of capturing such details in predictive models.

\begin{figure}[ht!]
\centering
\includegraphics[width=0.9\linewidth]{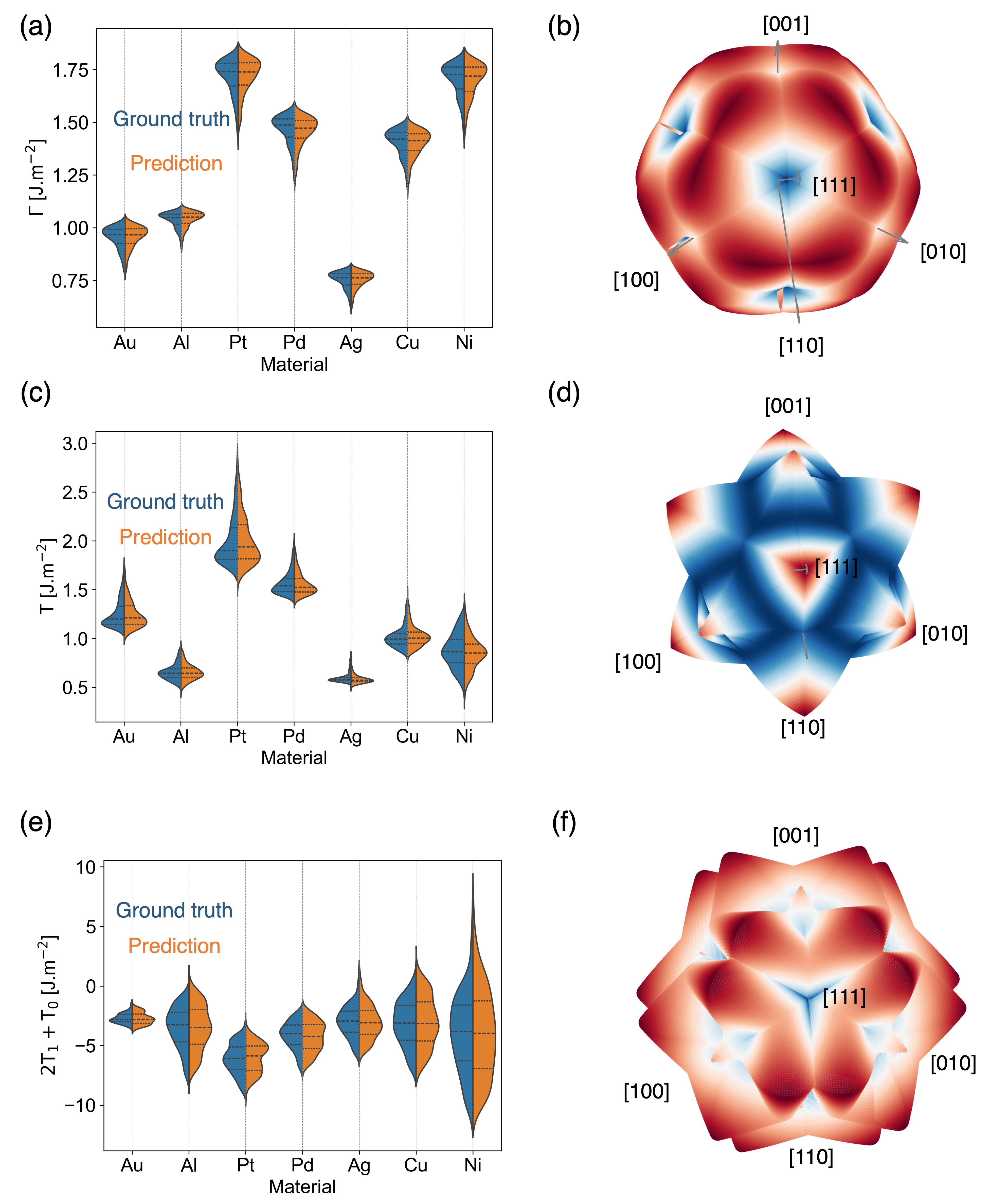}
\caption{\textbf{Visualizing the predictions of the nn-EQL:}
(a), (c), and (e) show distributions of the surface tension $\Gamma$, residual surface stress invariant $T$, and the isotropic surface elasticity stiffness invariant $2 T_1+T_0$, respectively, comparing results from the semi-analytical method (blue) and the equations learned by the nn-EQL (orange).
Surface plots in (b), (d), and (f) illustrate the orientation dependence of $\Gamma$, $T$, and $2 T_1+T_0$ in Au predicted by the nn-EQL. (For interpretation of the references to color in this figure legend, the reader is
referred to the web version of this article.)}\label{fig3}
\end{figure}

The nn-EQL not only fits these distributional features correctly but it also generalizes to the dependence of these surface properties over the surface orientation space.
As illustrated in Fig. \ref{fig3} (b), (d), and (f) for Au, the analysis of surface orientation through Miller indices (e.g., $\langle100\rangle$ and $\langle111\rangle$) and orientation dependence reveals notable physical trends.
Stable low-index surfaces (i.e. surfaces with the lowest surface tension values, e.g., $\langle111\rangle$ or $\langle 100 \rangle$) show minimum or maximum elastic properties (see for instance preferred directions for the residual surface stress invariant in panel (d)), dominating the tails of their property distributions, while higher-index, less stable surfaces correspond to transition regions between these stable surfaces.
For instance, in the case of Au, we observe that the stable $\langle 111 \rangle$ direction has low values for $\Gamma$ and $2 T_1+T_0$, while intermediate, high Miller index surfaces have high values for $\Gamma$ and $2 T_1+T_0$, indicating the unstable nature of these surfaces.
Across these trends, we find that the nn-EQL can capture these variations and insights well, discovering relationships between surface orientations and property of interest, for seven different FCC materials, and across five different surface properties.
}

\begin{figure}[ht!]
\centering
\includegraphics[width=0.99\linewidth]{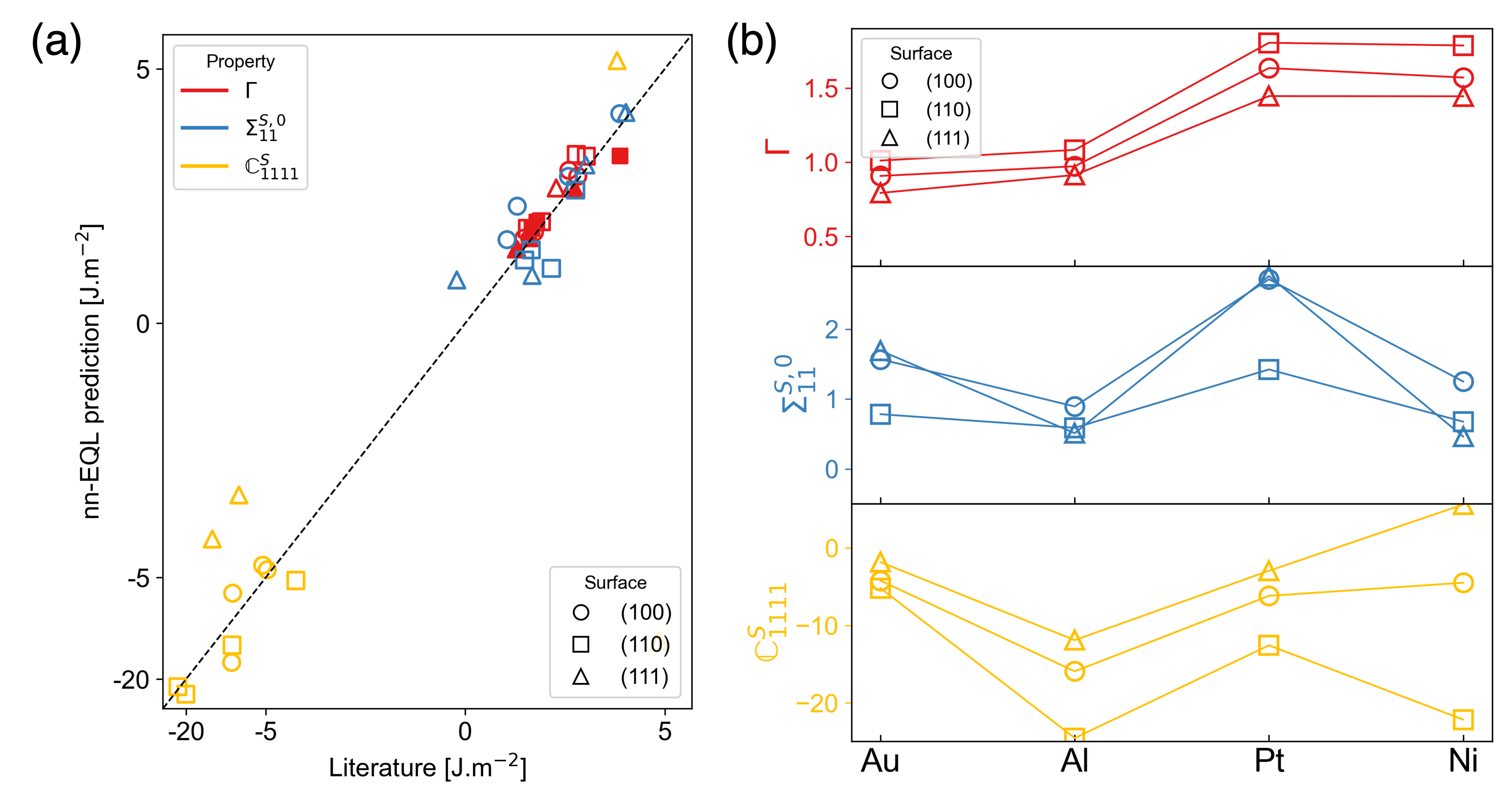}
\caption{\textbf{Comparing nn-EQL predictions to current literature:} (a) Comparing nn-EQL predicted surface energy/stress/elastic constants to literature \cite{shenoy2005atomistic} (molecular dynamics, open symbols), \cite{jain2013commentary} (density functional theory, filled symbols) (b) Trends in nn-EQL predicted surface properties with surface orientation and material.}\label{fig4}
\end{figure}

%\subsection{Comparison with existing models}
{
The nn-EQL predictions also display consistently good agreement with available surface elasticity data in the literature for low-index surfaces ($\langle 111 \rangle$, $\langle 110 \rangle$, $\langle 100 \rangle$) calculated with traditional simulation techniques like molecular dynamics \cite{shenoy2005atomistic} and density functional theory \cite{jain2013commentary}, see Fig. \ref{fig4} (a).
}

{
The discrepancies observed in Fig. \ref{fig4}(a) can be primarily attributed to differences in computational methodologies, specifically, the choice of interatomic potential used for molecular dynamics simulations~\cite{shenoy2005atomistic} or the underlying level of theory used in in the density functional theory calculations~\cite{jain2013commentary}.
These methodological choices directly influence the computed values of surface properties and often lead to variation between datasets.
The training data for the nn-EQL originates from Chen et al.~\cite{chen2022invariant}, where surface stresses and elastic constants were evaluated using a semi-analytical method, as opposed to explicit molecular dynamics simulations as in the case of Shenoy et al. \cite{shenoy2005atomistic}.
Additional sources of error may arise from finite-size and surface relaxation effects in atomistic simulations, as well as from other effects such as electronic effects as it would the case in density functional theory or whether surface reconstruction is accounted for in the modeling protocol.
We note that the magnitude of errors varies across surface properties: the largest discrepancies are associated with the surface elastic stiffness tensor components, followed by residual surface stress, while surface energies display the smallest erros.
This trend stems from the fundamental physical relationships among these quantities.
Indeed, as discussed in the following section through the interpretation of the functional relationships discovered by the nn-EQL, surface energy is predominantly governed by crystallography, resulting in robust and transferable predictions.
In contrast, residual surface stress and surface elastic stiffness are more sensitive to surface relaxation, reconstruction, and other orientation-dependent phenomena, making them more susceptible to methodological differences.
As shown in Fig. \ref{fig4}(b), the nn-EQL predictions capture material trends across low-index surfaces: $\langle 110\rangle$ surface energies are consistently higher than those of $\langle 100\rangle$ and $\langle 111\rangle$ for Au, Al, Pt, and Ni, whereas $\langle111\rangle$ surfaces generally exhibit the highest surface stiffness values.
Surface stress along the x-direction (component `11'), however, are found to be similar for the materials studied here.
This nuanced behavior, particularly for higher-order surface properties, highlights the sensitivity of these quantities to both crystallography and detailed atomistic configurations, as well as the importance of the modeling protocol used to extract them.
}

\subsection{Equation discovery for surface properties}

{
Following the above demonstration that the nn-EQL accurately predicts distributions of surface energies and key surface elastic invariants across broad material families, we now turn to the functional forms returned by the nn-EQL.
Starting with the surface tension $\Gamma$, the nn-EQL consistently predicts a seemingly universal, closed-form expression that captures orientation-dependent energetics common across the FCC metals tested.
This orientation-dependent functional form is expressed as:
\begin{equation}
\rm \Gamma^{\alpha} = \Gamma^\alpha_0 g_\Gamma\left(\theta,\phi\right) = \Gamma^\alpha_0\left(1 + A_0\theta + B_0 \phi + C_0 \sin(A_1 \theta - B_1 \phi)\right),
\label{eq:gamma_function}
\end{equation}
for $\alpha =\{\textrm{Au, Al, Pt, Pd, Ag, Cu, Ni}\}$.
The term $\Gamma^\alpha_0$ represents an intrinsic, baseline, orientation-independent surface energy value which correlates with atomic species, while the dimensionless function $\rm g_\Gamma\left(\theta,\phi\right)$ encodes the variation of surface energy with surface orientation.
Examination of the coefficients of $g_\Gamma\left(\theta,\phi\right)$ immediately reveals the systematic underlying crystal-specific constraints.
Moreover, we note that coefficients $B_0$ and $C_0$ are similar for most materials entries, indicating a common scaling in the linear and periodic orientation contributions to the surface energy.
Values of the various coefficients are listed in Table~\ref{tab:coeffs_gamma}.
\begin{table}[!ht]
\centering
\caption{Coefficients in the general expression for $\Gamma$ for various FCC materials}
\begin{tabular}{|c|c|c|c|c|c|c|}
\hline
\hline
& \rm{$\Gamma_0$} [J/m$^2$] & \textbf{$A_0$} & \textbf{$B_0$} & \textbf{$C_0$} & \textbf{$A_1$} & \textbf{$B_1$}\\
\hline
{\bf Au} & 0.62 & 0.10 & 0.31 & 0.37 & 1.37 & 1.87\\
\hline
{\bf Al} & 0.80 & 0.07 & 0.05 & 0.22 & 1.68 & 1.83\\
\hline
{\bf Pt} & 1.19 & 0.13 & 0.11 & 0.27 & 1.57 & 1.89\\
\hline
{\bf Pd} & 1.00 & 0.11 & 0.11 & 0.32 & 1.57 & 1.75\\
\hline
{\bf Ag} & 0.52 & 0.12 & 0.15 & 0.29 & 1.62 & 2.14\\
\hline
{\bf Cu} & 1.01 & 0.11 & 0.12 & 0.24 & 1.70 & 2.13\\
\hline
{\bf Ni} & 1.24 & 0.08 & 0.12 & 0.26 & 1.65 & 2.02\\
\hline
\hline
\end{tabular}
\label{tab:coeffs_gamma}
\end{table}
}

{
The same principle applies to higher-order invariants, like the residual surface stress invariant, $T$ for instance.
Here again, the nn-EQL reveals a generalized closed-form expression that capture orientation-dependent of the residual surface stress invariant such that,
\begin{equation}
\begin{split}
    T_\alpha & = T^\alpha_0 g_T(\theta,\phi) \\
    & = T^\alpha_0\Big(1 + A_0 \theta + B_0 \phi + C_0\, \sin(A_1 \theta - B_1 \phi)
    + C_1\, \sin(A_2 \theta + B_2 \phi)\Big),
\end{split}
\label{eq:T_function}
\end{equation}
}
\noindent for $\alpha =\{\textrm{Au, Al, Pt, Pd, Ag, Cu, Ni}\}$.
In the equation above, the term $T^\alpha_0$ denotes an intrinsic residual surface stress.
However, the variation of residual surface stress with respect to the surface orientation suggests a more complex variation, possibly linked to complex surface structures, surface relaxation effects, or surface reconstructions for high-index surfaces~\cite{bach1997stress}.
Table~\ref{tab:coeffs_T} lists the various coefficients associated with the functional for all the FCC metals tested.
The full set of equations for each surface elastic invariant and for each materials is provided in the Appendix.

\begin{table}[!ht]
\centering
\caption{Coefficients in the general expression for $T$ for various materials}
\begin{tabular}{|c|c|c|c|c|c|c|c|c|c|}
\hline
\hline
& \textbf{$T_0$} [J/m$^2$] & \textbf{$A_0$}   & \textbf{$B_0$} & \textbf{$C_0$} & \textbf{$A_1$} & \textbf{$B_1$} & \textbf{$C_1$} & \textbf{$A_2$} & \textbf{$B_2$}\\
\hline
{\bf Au} & 1.82 & 0.13 & -0.17 & -0.50 & 1.53 & 1.73 & 0 & 0 & 0\\
\hline
{\bf Al} & 0.40 & 0.37 & 0.14 & -1.48 & 0.22 & -1.55 & 1.48 & 0.57 & 0.64\\
\hline
{\bf Pt} & 2.10 & 0.51 & 0.20 & -0.77 & 1.62 & 1.63 & 0.30 & 2.19 & 1.73\\
\hline
{\bf Pd} & 1.67 & 0.54 & -0.24 & -0.59 & 1.35 & 1.52 & 0.26 & 2.39 & 1.87\\
\hline
{\bf Ag} & 0.45 & 0.24 & -0.07 & -1.09 & 0.38 & -2.92 & 1.14 & 0.72 & 2.33\\
\hline
{\bf Cu} & 0.51 & 0.82 & -0.22 & -1.12 & 0.30 & -2.53 & 1.04 & 0.72 & 1.61\\
\hline
{\bf Ni} & 0.20 & 2.22 & -2.80 & 2.12 & 0.69 & -0.21 & 0 & 0 & 0\\
\hline
\hline
\end{tabular}
\label{tab:coeffs_T}
\end{table}

{
The analytic forms discovered by the nn-EQL do not simply interpolate over the training data, they highlight symmetry constraints and element-specific behaviors.
Indeed, a closer examination of the orientation dependence of the surface tension $\Gamma$ reveals striking similarity across all the materials studied.
The function $g_\Gamma(\theta,\phi)$ consistently defines a plane $A_1\theta - B_1\phi = \pi/2 + 2\pi k$ ($k = [0, 1]$) in the $(\theta,\phi)$ orientation space corresponding to a direction of maximum surface energy.
This universal behavior suggests that despite differences in specific atomic species and baseline surface energy magnitudes ($\Gamma_0^\alpha$), the geometric constraints and crystallographic symmetries of the FCC lattice impose nearly invariant angular pattern for the surface energy variation with respect to the surface orientation.
}

{
In contrast, for the residual surface stress invariant $T$, the orientation dependence $g_T\left(\theta,\phi\right)$ is more complex and material specific, as indicated by the non-zero $C_1,A_2,B_2$ coefficients in some materials.
Unlike $\Gamma$, $T$ does not follow a universal orientation trend across the FCC metals surveyed.
Rather, certain elements (Al, Pt, Pd, Ag, and Cu) feature an additional sinusoidal that further accentuates anisotropic variations.
Even for the other two elements (Au, Ni) that do not display this added anisotropic behavior, their orientation dependence defers from that of their corresponding $\Gamma$ functional form.
This analysis, comparing the orientation dependence of $\Gamma$ and $T$, reveals that surface relaxation, surface reconstruction, and electronic structure play intricate roles in determining $T$~\cite{needs1987calculations, needs1991theory, ibach1997role, shenoy2005atomistic, dingreville2007semi}, but their influence is less dominant than that of crystallographic atomic packing, which primarily governs the surface tension $\Gamma$.
In other words, these equations reveal that the residual surface stress is highly sensitive to variations in local atomic environments and bonding, extending beyond what can be explained by crystallographic factors alone.
Thus, the nn-EQL not only provides physically interpretable forms for orientation dependence on surface elastic constant, but it also reveals richer complexities for certain surface properties emphasizing the intricate interplay between crystal symmetry, and relaxation effects dictating surface mechanical behaviors.
}

\subsection{Correlating surface properties to bulk properties}
{
The functional equations presented in the section above successfully decoupled the complex, orientation-dependent of geometry properties (e.g., $g_\Gamma(\theta,\phi)$ and $g_T(\theta,\phi)$ functions) from their materials-specific scaling coefficients (e.g., $\Gamma^{\alpha}_0$,$ T^{\alpha}_0$, $R^{\alpha}_0$, etc.).
These coefficients, which capture the unique ``signature'' of each metal, must logically correspond to the material's fundamental bulk properties.
In this section, we test this hypothesis by seeking direct, quantitative correlations between these surface coefficients and a physically motivated set of bulk properties. 
}

\begin{figure}[h!]
\centering
\includegraphics[width=0.99\linewidth]{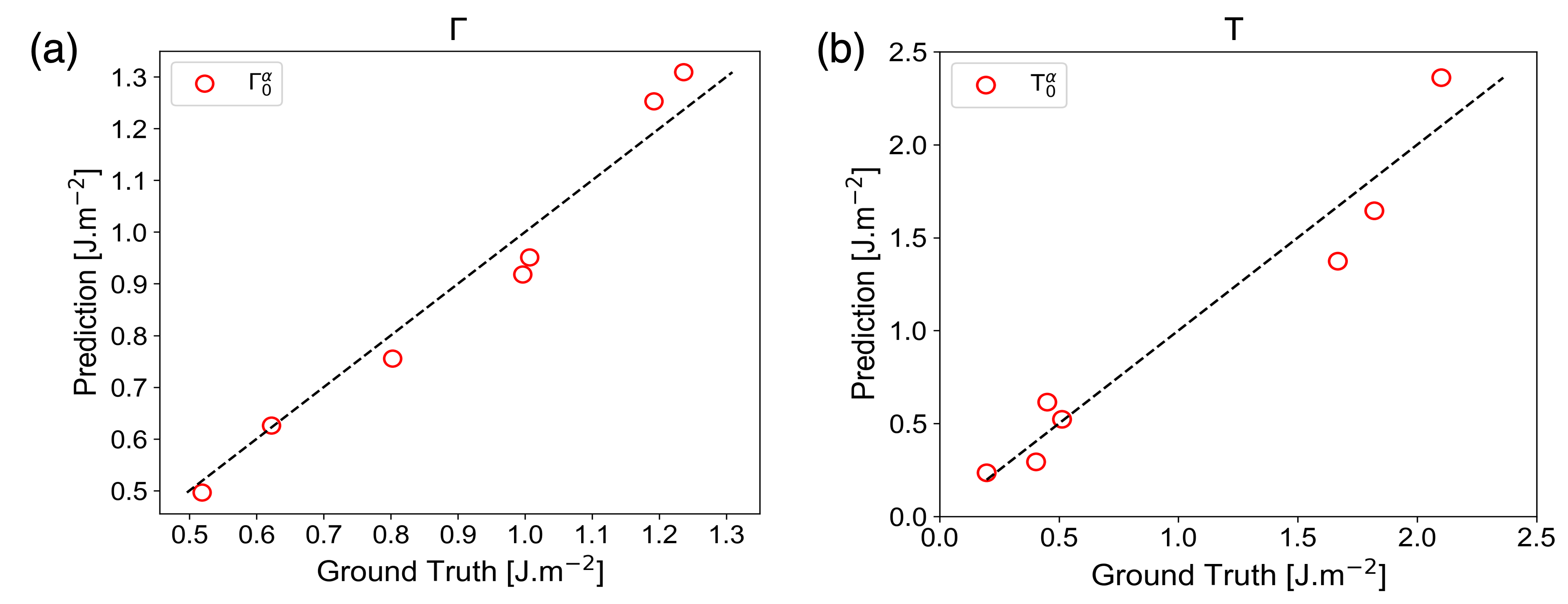}
\caption{\textbf{Parity plots of the material-specific coefficients in the equations for the surface invariants.} `Ground Truth' values are from the nn-EQL discovery and `Predictions' values are from the linear model based on normalized bulk properties from Eq\@.~\eqref{eq:linmodel}. Plots correspond to coefficients for (a) $\Gamma$ and (b) $T$.}\label{fig5}
\end{figure}

{
To establish this relationship, we selected bulk properties that are directly related to the elastic response of bulk crystals.
Because the surface energy pertains to the energy required to create free surfaces, it is reasonable to consider the cohesive energy $E_{\rm{coh}}$, which represents the total energy required to separate all atoms in the crystal, as a correlation descriptor.
Similarly, the stacking-fault energy $E_{\rm{SF}}$ is another energy analog,
representing the energy cost of creating a 2D planar defect, and sharing a similar physical origin to surface energy.
We also consider the fundamental elastic moduli: the shear moduli $G' = \frac{1}{2}(\mathbb{C}_{1111} - \mathbb{C}_{1122})$ and $G'' = \mathbb{C}_{2323}$ and the bulk modulus $K$.
To establish a physically meaningful relationship and ensure dimensional consistency, we further normalized these bulk properties to have units of energy per unit area (J/m$^2$), matching the units of the surface properties.
The stacking-fault energy, $E_{\rm{SF}}$, is already in J/m$^2$.
The cohesive energy, $E_{\rm{coh}}$, is normalized by a characteristic atomic area, $a^2$, where $a$ is the material specific lattice parameter, to yield $E_{\rm{coh}}/a^2$ in J/m$^2$.
The elastic moduli ($G'$, $G''$, $K$), all in J/m$^3$, are multiplied by $a$ to yield $aG'$, $aG''$, $aK$ all in J/m$^2$.
Therefore, we are not simply fitting abstract coefficients.
We are testing the physical hypothesis that the material-specific components of surface elasticity can be expressed as a linear combination of the material's other fundamental bulk properties.
Given our small dataset of seven materials, we seek the simplest form of this relationship, a linear model, to capture the primary physical trends without overfitting.
As such a generic form for this linear model can be expressed as:
}
\begin{equation}
{\rm y^\alpha} = w_{G'}\cdot aG' + w_{G''}\cdot aG'' + w_{K}\cdot aK + w_{\rm{SF}}\cdot E_{\rm{SF}} + w_{\rm{coh}}\cdot E_{\rm{coh}}/a^2,\label{eq:linmodel}
\end{equation}
{
\noindent where $\rm y^\alpha$ is the set of coefficients describing the equation for a given property, for material $\alpha$, i.e., ${\rm y^\alpha} = \{\Gamma^{\alpha}_0, T^{\alpha}_0, R^{\alpha}_0, ...\}$.
As shown in the parity plots in Fig\@.~\ref{fig5}, we find that a linear relationship provides an accurate description of the surface elastic constants in terms of the bulk materials properties.
}

{
The results of this linear fit for the baseline, orientation-independent term, $\Gamma_0^\alpha$ and $T_0^\alpha$, are shown in Table~\ref{tab:bulk_coeff1}.
The analysis of these coefficients provides confirmation on physical correlations.
For $\Gamma_0^\alpha$, the weighting coefficient for $E_{\rm{SF}}$ ($w_{\rm{SF}}$) is at least an order of magnitude larger than any other coefficient.
After accounting for the scale of $E_{\rm{SF}}$, we find that the stacking fault energy still plays a critical role in determining the weighting coefficients.
This is in agreement with the sensitivity analysis provided in~\cite{chen2022invariant}, this indicates that $E_{\rm{SF}}$ is the single most dominant predictor.
The negative sign reveals a strong inverse correlation: materials with higher stacking-fault energies tend to have lower baseline surface energies.
In contrast, the influences of the normalized shear moduli ($w_{G'}$) and cohesive energy ($w_{\rm{coh}}$) are secondary, while the correlation with $w_{G''}$ is quite strong.
The correlation bulk modulus ($w_{K}$) is negligible for $\Gamma^{\alpha}_0$, even after accounting for the numerical scale of the bulk modulus for these materials.
An even stronger correlation is observed for the residual stress coefficient, $T_0^\alpha$.
In this case, the stacking-fault energy is again a dominant factor, but with a positive correlation.
This implies that a high stacking-fault energy strongly correlates with a high (tensile) residual surface stress.
The normalized shear moduli and cohesive energy once have a moderate, non-negligible influence on the residual surface stress.
However, in this case the bulk modulus also plays a strong role in determining the weighting coefficient.
In both cases, the stacking-fault energy is conclusively the most critical bulk parameter correlated with surface properties, confirming trends identified in previous sensitivity analysis (see Fig\@.~15 in~\cite{chen2022invariant}).
A complete list of coefficients for all surface invariants is provided in the Appendix.
}

\begin{table}[!ht]
\centering
\caption{Coefficients relating surface properties to bulk properties}
\begin{tabular}{|c|c|c|c|c|c|}
\hline
\hline
& $w_{G'}$ & $w_{G''}$ & $w_{K}$ & $w_{\rm{SF}}$ & $w_{\rm{coh}}$\\
\hline
$\Gamma^{\alpha}_0$ & -0.288 & 0.120 & -0.004 & -7.65 & -0.052\\
\hline
$T^{\alpha}_0$ & 0.495 & -0.234 & 0.038 &  20.23 & -0.195\\
\hline
\hline
\end{tabular}
\label{tab:bulk_coeff1}
\end{table}
%

%
%%%%%%%%%%%%%%%%%%%%%%%%%%%%%%%%%%%%%%%%%%%%%%%%%%%%%%%%%%%
%% CONCLUSION
%%%%%%%%%%%%%%%%%%%%%%%%%%%%%%%%%%%%%%%%%%%%%%%%%%%%%%%%%%%
\section{Conclusion}\label{sec:conclusion}
{
This work demonstrates that a neural network equation learner can serve as a powerful approach for discovering interpretable, closed-form relationships connecting bulk properties to surface elastic properties.
The model, which utilizes customized activation functions and a connection-based pruning strategy to yield compact symbolic equations, accurately predicts orientation-dependent surface elastic invariants across seven FCC metals, effectively capturing complex phenomena such as long-tailed distributions arising from high-index surfaces.
Additionally, the learned analytical expressions separate surface properties into generalized orientation-driven functions and material-dependent baseline coefficients, providing direct physical interpretablity.
}

{
Analysis of these equations delivered physical insights on the dependence of surface properties.
For instance, the orientation dependence for surface tension was found to be nearly universal across all seven FCC metals, strongly suggesting that surface energy is governed primarily by the geometric and crystallographic constraints of the FCC lattice.
In contrast, the orientation dependence for residual surface stress was found to be highly material-specific, connected to other phenomena like surface relaxation, reconstruction, and local atomic bonding that go beyond simple crystallographic factors.
These findings establish a robust, quantitative link between bulk and surface elastic properties.
}

{
Nevertheless, several limitations must be emphasized.
The predictive equations are trained on atomistic data derived from empirical potentials, which may omit important physical effects such as electronic structure contributions to surface relaxation for instance~\cite{nichols2002surface, kadas2006surface}.
It remains an open question whether the discovered equations would persist if trained on more physically accurate data (e.g. DFT).
This ``missing physics'' is for instance apparent in our choice of bulk properties.
Thus, certain key bulk descriptors, such as Fermi energy or other electronic properties, are not present in the current model, and their inclusion could significantly enhance the generality and transferability of the equations to a broader materials space.
A valuable next step is to integrate higher-fidelity reference data (e.g., from DFT), incorporate electronic descriptors, and systematically test the discovered relationships across more diverse materials systems.
Similarly, on a methodological perspective, the data used to train the equation learner originated from noiseless semi-analytical data.
It is an open question whether the discovered equations would change if fit to noisy, experimental data, or data derived from other measurements with uncertainties.
}

{
As a last point, these results illustrate an approach beyond ``black-box'' machine learning by using interpretable symbolic regression neural network to extract physically meaningful equations.
The methodology outlined here enables direct embedding of learned equations as symbolic constitutive laws for continuum and multiscale models.
This opens a pathway to predicting and understanding the mechanical behavior of nanostructured materials, whose size-dependent properties arise from the interplay between bulk and surface properties~\cite{dingreville2005surface, ansari2011bending, li2014dependence}.
Ultimately, this interpretable equation learning approach enriches the toolkit for bridging atomistic detail and continuum-level predictions to accelerate discovery in the mechanics of complex materials.
}
\section*{CRediT authorship contribution statement}
% {\bf S. Desai:} Methodology/Study design, Software, Validation, Formal analysis, Investigation, Data curation, Writing – original draft, Writing – review and editing, Visualization.
% {\bf P. Iyer:} Methodology/Study design, Investigation,  Writing – original draft, Writing – review and editing.
% {\bf R. Dingreville:} Conceptualization, Methodology/Study design, Software, Validation, Formal analysis, Investigation, Resources, Writing – original draft, Writing – review and editing, Visualization, Supervision, Funding acquisition.
%%%%%%%%%%%%%%%%%%%%%%%%%%%%%%
%% ACKNOWLEDGEMENT
%%%%%%%%%%%%%%%%%%%%%%%%%%%%%%
\section*{Acknowledgments}
{
\par{
The authors would like to thank Prof\@.~Krishna Garikipati and Jingye Tan from the University of Southern California and Philip Jacobson from Sandia National Laboratories for comments and suggestions on the initial draft of this manuscript.
The authors are supported by the U.S\@.~Department of Energy, Office of Science, Office of Advanced Scientific Computing Research and  Office of Basic Energy Sciences, Scientific Discovery through Advanced Computing (SciDAC) program under the MIRAGE project.
The atomistic simulations used computational resources provided by the Center for Integrated Nanotechnologies (CINT), an Office of Science user facility operated for the U.S\@.~Department of Energy.
This article has been authored by an employee of National Technology \& Engineering Solutions of Sandia, LLC under Contract No\@.~DE-NA0003525 with the U.S\@.~Department of Energy (DOE).
The employee owns all right, title, and interest in and to the article and is solely responsible for its contents. The United States Government retains and the publisher, by accepting the article for publication, acknowledges that the United States Government retains a non-exclusive, paid-up, irrevocable, world-wide license to publish or reproduce the published form of this article or allow others to do so, for United States Government purposes.
The DOE will provide public access to these results of federally sponsored research in accordance with the DOE Public Access Plan https://www.energy.gov/downloads/doe-public-access-plan. 
}
}

\section*{Data availability}
\par{
Data will be made available upon request.
}
\section*{Declaration of competing interests}
\par{
There are no competing interests to declare.
}
% 

\appendix
\setcounter{figure}{0}
\setcounter{table}{0}
\section{Functional forms for all surface property invariants}
{
This Appendix lists the complete functional forms discovered for each surface invariant for the seven FCC materials analyzed in this study.
}
\subsection*{Surface Tension, $\Gamma$}
\begin{equation}
\begin{aligned}
\Gamma^{{\rm Au}} &= 0.62\times(1 + 0.10 \theta + 0.31 \phi + 0.37 \rm sin(1.37 \theta - 1.88 \phi)) \\
\Gamma^{{\rm Al}} &= 0.80\times(1 + 0.07 \theta + 0.05 \phi + 0.22 \rm sin(1.68 \theta - 1.83 \phi)) \\
\Gamma^{{\rm Pt}} &= 1.19\times(1+ 0.13 \theta + 0.11 \phi + 0.27 \rm sin(1.57 \theta - 1.89 \phi)) \\
\Gamma^{{\rm Pd}} &= 1.00\times(1 + 0.11 \theta + 0.11 \phi + 0.32 \rm sin(1.57 \theta - 1.75 \phi) ) \\
\Gamma^{{\rm Ag}} &= 0.52\times(1+ 0.12 \theta + 0.15 \phi + 0.29 \rm sin(1.62 \theta - 2.13 \phi)) \\
\Gamma^{{\rm Cu}} &= 1.01\times(1+ 0.11 \theta + 0.12 \phi + 0.24 \rm sin(1.70 \theta - 2.13 \phi)) \\
\Gamma^{{\rm Ni}} &= 1.24\times(1+ 0.08 \theta + 0.12 \phi + 0.26 \rm sin(1.65 \theta - 2.02 \phi))
\label{eq:surfaceprops_Gamma0}
\end{aligned}
\end{equation}

\begin{equation}
\begin{aligned}
\Gamma^{\alpha} &= \Gamma_0^\alpha(1+ A_0 \theta + B_0 \phi + C_0 \rm sin(A_1 \theta - B_1 \phi))
\label{eq:surfaceprops_Gamma0_general}
\end{aligned}
\end{equation}

\subsection*{Residual Surface Stress, $\Sigma^{S,0}$}
\begin{equation}
\begin{aligned}
T^{{\rm Au}} &= 1.82\times(1 +0.13 \theta - 0.17 \phi - 0.50 \rm sin(1.53 \theta + 1.73 \phi)) \\
T^{{\rm Al}} &= 0.40\times(1 +0.37 \theta + 0.14 \phi - 1.48 \rm sin(0.22 \theta - 1.55 \phi) + 1.48 \rm sin(0.57 \theta + 0.64 \phi)) \\
T^{{\rm Pt}} &= 2.10\times(1 +0.51 \theta + 0.20 \phi - 0.77 \rm sin(1.62 \theta + 1.63 \phi) + 0.30 \rm sin(2.19 \theta + 1.73 \phi)) \\
T^{{\rm Pd}} &= 1.67\times(1 +0.54 \theta - 0.24 \phi - 0.59 \rm sin(1.35 \theta + 1.52 \phi) + 0.26 \rm sin(2.39 \theta + 1.87 \phi)) \\
T^{{\rm Ag}} &= 0.45\times(1 +0.24 \theta - 0.07 \phi - 1.09 \rm sin(0.38 \theta - 2.92 \phi) + 1.14 \rm sin(0.72 \theta + 2.33 \phi)) \\
T^{{\rm Cu}} &= 0.51\times(1 +0.82 \theta - 0.22 \phi - 1.12 \rm sin(0.30 \theta - 2.53 \phi) + 1.04 \rm sin(0.72 \theta + 1.61 \phi)) \\
T^{{\rm Ni}} &=  0.20\times(1 + 2.22 \theta - 2.80 \phi + 2.12 \rm sin(0.69 \theta - 0.21 \phi)) \\
\label{eq:surfaceprops_T}
\end{aligned}
\end{equation}

\begin{equation}
\begin{aligned}
T^{\alpha} &= T_0^\alpha(1 + A_0 \theta + B_0 \phi + C_0 \rm sin(A_1 \theta - B_1 \phi) + C_1 \rm sin(A_2 \theta + B_2 \phi)) \\
\label{eq:surfaceprops_T_general}
\end{aligned}
\end{equation}

\begin{equation}
\begin{aligned}
R^{{\rm Au}} &= -0.19\times(1 + 0.90 \theta - 4.18 \phi + 2.85 \rm sin(0.10 \theta - 2.46 \phi) - 3.83 \rm sin(1.43 \theta - 1.31 \phi)) \\
R^{{\rm Al}} &= 0.01\times(1 - 9.30 \theta + 11.25 \phi - 7.25 \rm sin(0.27 \theta - 1.98 \phi) + 19.7 \rm sin(1.34 \theta - 1.21 \phi)) \\
R^{{\rm Pt}} &= -0.30\times(1 + 2.96 \theta - 8.11 \phi - 3.45 \rm sin(0.23 \theta + 2.80 \phi) - 5.15 \rm sin(1.34 \theta - 1.21 \phi))\\
R^{{\rm Pd}} &= - 0.09\times(1 + 5.59 \theta - 10.46 \phi - 6.02 \rm sin(0.26 \theta + 2.94 \phi) - 10.55 \rm sin(1.49 \theta - 1.34 \phi)) \\
R^{{\rm Ag}} &= 0.13\times(1 - 3.39 \theta + 6.75 \phi + 4.43 \rm sin(0.11 \theta + 2.17 \phi) + 4.53 \rm sin(1.42 \theta - 1.31 \phi)\\
R^{{\rm Cu}} &= 0.50\times(1 - 0.15 \theta + 0.87 \phi + 0.93 \rm sin(0.29 \theta - 3.47 \phi) - 1.84 \rm sin(0.45 \theta + 2.05 \phi))\\
R^{{\rm Ni}} &= 0.25\times(1 + 0.35 \theta + 0.12 \phi - 2.28 \rm sin(0.95 \theta - 0.27 \phi) + 1.30 \rm sin(1.62 \theta - 1.22 \phi)) \\
\label{eq:surfaceprops_R}
\end{aligned}
\end{equation}

\begin{equation}
\begin{aligned}
R^{\alpha} &= R_0^\alpha(1 + A_0 \theta + B_0 \phi + C_0 \rm sin(A_1 \theta - B_1 \phi) + C_1 \rm sin(A_2 \theta + B_2 \phi))\\
\label{eq:surfaceprops_R_general}
\end{aligned}
\end{equation}

\subsection*{Surface Elasticity Stiffness, $\mathbb{C}^{S}$}
\begin{equation}
\begin{aligned}
T_{0}^{\rm Au} &= -1.24 \times (1 + 0.94\theta -0.65 \phi -0.81 \rm sin(0.55\theta + 2.13\phi))\\
T_{0}^{\rm Al} &= -2.97 \times (1 + 0.03\theta + 1.01 \phi -0.10 \rm sin(0.18\theta + 0.15\phi) -0.70 \rm sin(1.84\theta + 1.56\phi))\\
T_{0}^{\rm Pt} &= -3.57 \times (1 + 0.56\theta -0.60 \phi -0.67 \rm sin(0.98\theta + 1.35\phi))\\
T_{0}^{\rm Pd} &= -2.39 \times (1 + 0.72\theta -0.61 \phi -0.90 \rm sin(1.12\theta + 1.08\phi))\\
T_{0}^{\rm Ag} &= -1.89 \times (1 + 0.53\theta -1.23 \phi -0.32 \rm sin(0.08\theta + 0.08\phi) -0.77 \rm sin(1.64\theta + 1.38\phi))\\
T_{0}^{\rm Cu} &= -3.31 \times (1 + 0.08\theta -1.22 \phi -0.05 \rm sin(0.17\theta + 0.14\phi) -0.86 \rm sin(1.85\theta + 1.50\phi))\\
T_{0}^{\rm Ni} &= -7.28 \times (1 -0.07\theta -1.11 \phi -0.09 \rm sin(0.18\theta + 0.14\phi) -0.76 \rm sin(1.87\theta + 1.50\phi))\\
\label{eq:surfaceprops_T0}
\end{aligned}
\end{equation}

\begin{equation}
\begin{aligned}
T_{0}^\alpha &= T_{0,0}^\alpha(1 + A_0 \theta + B_0 \phi + C_0 \rm sin(A_1 \theta - B_1 \phi) + C_1 \rm sin(A_2 \theta + B_2 \phi)) \\
\label{eq:surfaceprops_T0_general}
\end{aligned}
\end{equation}

\begin{equation}
\begin{aligned}
T_{1}^{\rm Au} &= - 0.83 \times (1 + 1.42 \theta - 5.00 \phi - 2.49 \rm sin(0.66 \theta - 3.23 \phi) - 1.13 \rm sin(1.52 \theta - 1.28 \phi))\\
T_{1}^{\rm Al} &= - 2.78 \times (1 + 0.04 \theta + 0.97 \phi - 1.39 \rm sin(\theta - 1.19 \phi)) \\
T_{1}^{\rm Pt} &= 2.21 \times (1 - 1.02 \theta - 0.22 \phi - 0.54 \rm sin(3.19 \theta - 3.41 \phi)) \\
T_{1}^{\rm Pd} &= -1.8 \times (-0.26 \theta + 1.12 \phi - 0.90 \rm sin(1.24 \theta + 1.13 \phi)) \\
T_{1}^{\rm Ag} &= -0.97 \times (1 + 0.80 \theta - 2.23 \phi - 2.48 \rm sin(1.75 \theta - 0.55 \phi) + 2.01 \rm sin(2.77 \theta - 2.44 \phi)) \\
T_{1}^{\rm Cu} &= 7.88 \times (1 + 0.04 \theta - 0.63 \phi - 0.86 \rm sin(1.21 \theta + 0.58 \phi)) \\
T_{1}^{\rm Ni} &= - 1.01 \times (1 + 1.85 \theta - 7.43 \phi - 5.11 \rm sin(1.70 \theta - 1.20 \phi) + 4.25 \rm sin(2.78 \theta - 2.42 \phi)) \\
\label{eq:surfaceprops_T1}
\end{aligned}
\end{equation}

\begin{equation}
\begin{aligned}
T_{1}^\alpha &= T_{1,0}^\alpha(1 + A_0 \theta + B_0 \phi + C_0 \rm sin(A_1 \theta + B_1 \phi) + C_1 \rm sin(A_2 \theta + B_2 \phi)) \\
\label{eq:surfaceprops_T1_general}
\end{aligned}
\end{equation}

\begin{equation}
\begin{aligned}
R_{0}^{\rm Au} &= 2.29 \times (1 + 0.16\theta -3.01\phi - 1.24\rm sin(0.47 \theta -2.54 \phi)) \\
R_{0}^{\rm Al} &= 0.27 \times (1 + 5.16 \theta -11.52\phi -5.36\rm sin(0.44 \theta + 4.01 \phi) + 2.47\rm sin(2.67 \theta -3.09 \phi)) \\
R_{0}^{\rm Pt} &= 1.76 \times (1 + 0.03 \theta -1.45\phi + 0.68\rm sin(2.37 \theta -2.68 \phi)) \\
R_{0}^{\rm Pd} &= 0.97 \times (1 - 0.10\theta -1.09 \phi + 0.88\rm sin(2.42 \theta -3.11 \phi)) \\
R_{0}^{\rm Ag} &= -0.06 \times (1 - 15.68\theta -1.38\phi + 18.34 \rm sin(0.47  \theta + 2.49 \phi) -7.77 \rm sin(3.12 \theta -4.67 \phi)) \\
R_{0}^{\rm Cu} &= -0.08 \times (1 - 27.26 \theta + 38.24 \phi + 22.44 \rm sin(0.41 \theta + 3.39 \phi) -9.81  \rm sin(2.89 \theta -3.03 \phi)) \\
R_{0}^{\rm Ni} &= 0.90 \times (1 + 5.52 \theta -11.58 \phi - 5.46 \rm sin(0.61  \theta - 3.32 \phi) + 2.59 \rm sin(2.47 \theta - 1.63 \phi)) \\
\label{eq:surfaceprops_R0}
\end{aligned}
\end{equation}

\begin{equation}
\begin{aligned}
R_{0}^\alpha &= R_{0,0}^\alpha(1 + A_0 \theta + B_0 \phi + C_0 \rm sin(A_1 \theta + B_1 \phi) + C_1 \rm sin(A_2 \theta + B_2 \phi)) \\
\label{eq:surfaceprops_R0_general}
\end{aligned}
\end{equation}

\begin{equation}
\begin{aligned}
R_{1}^{\rm Au} &= -0.55 \times (1 - 1.42\theta + 0.56 \phi + 1.67 \rm sin(0.59 \theta + 2.22 \phi) - 1.53 \rm sin(1.87 \theta - 0.55 \phi)) \\
R_{1}^{\rm Al} &= 5.38 \times (1 + 0.20 \theta + 2.13 \phi - 3.06 \rm sin(0.04  \theta + 2.03 \phi) - 1.03 \rm sin(0.31 \theta - 3.57 \phi)) \\
R_{1}^{\rm Pt} &= -0.30 \times (1 - 3.93 \theta - 1.84 \phi + 6.89 \rm sin(0.26  \theta + 1.57 \phi) - 3.40 \rm sin(2.45 \theta - 1.90 \phi)) \\
R_{1}^{\rm Pd} &= 0.12 \times (1 + 7.14 \theta - 11.46 \phi - 5.09 \rm sin(1.16  \theta + 0.54 \phi) + 6.83 \rm sin(2.81 \theta - 2.21 \phi)) \\
R_{1}^{\rm Ag} &= 3.25 \times (1 - 0.23 \theta + 1.07 \phi - 2.20 \rm sin(0.02 \theta + 2.48 \phi) - 1.05 \rm sin(0.33 \theta - 3.82 \phi)) \\
R_{1}^{\rm Cu} &= 5.59 \times (1 - 0.28\theta + 1.50 \phi + 2.29 \rm sin(0.02  \theta - 2.45 \phi) - 0.90 \rm sin(0.45 \theta - 4.11 \phi)) \\
R_{1}^{\rm Ni} &= 13.73 \times (1 - 0.03 \theta + 2.88 \phi - 4.0 \rm sin(0.10 \theta + 2.03 \phi) - 1.35 \rm sin(0.15 \theta - 3.42 \phi)) \\
\label{eq:surfaceprops_R1}
\end{aligned}
\end{equation}

\begin{equation}
\begin{aligned}
R_{1}^\alpha &= R_{1,0}^\alpha(1 + A_0 \theta + B_0 \phi + C_0 \rm sin(A_1 \theta + B_1 \phi) + C_1 \rm sin(A_2 \theta + B_2 \phi)) \\
\label{eq:surfaceprops_R1_general}
\end{aligned}
\end{equation}

\section{Correlation of surface invariant properties to bulk properties}

Table \ref{tab:bulk_coeff2} documents the rest of the coefficients describing how the material-specific coefficients depend on the bulk properties of the material.

\begin{table}[!ht]
\centering
\caption{Coefficients relating surface properties to bulk properties}
\begin{tabular}{|c|c|c|c|c|c|}
\hline
\hline
& $w_{G'}$ & $w_{G''}$ & $w_{K}$ & $w_{\rm{SF}}$ & $w_{\rm{coh}}$\\
\hline
$R^{\alpha}$ & 0.215 & -0.086 & -0.012 & 23.18 & -0.292\\
\hline  
$R^{\alpha}_{0,0}$ & 1.052 & -0.469 & 0.030 & 29.88 & -0.754\\
\hline   
$R^{\alpha}_{1,0}$ & -1.979 & 1.101 & -0.206 & -126.4 & -0.951\\
\hline
$T^{\alpha}_{0,0}$ & 2.033 & -0.938 & 0.056 & 80.95 & -0.447\\
\hline 
$T^{\alpha}_{1,0}$ & 5.131 & -2.540 & -0.013 & 610.7 & -7.448\\

\hline
\hline
\end{tabular}
\label{tab:bulk_coeff2}
\end{table}
% 

\bibliographystyle{elsarticle-num} 
\bibliography{sn-bibliography.bib}

\end{document}